\documentstyle[prb,aps,epsfig,floats]{revtex}

\newcommand{\vc}[1]{{\bf #1}}

\begin{document}

\draft
\twocolumn[\hsize\textwidth\columnwidth\hsize\csname
@twocolumnfalse\endcsname

\title{Ab initio study of BaTiO$_3$ and PbTiO$_3$ surfaces in external
       electric fields}

\author{B. Meyer and David Vanderbilt }

\address{Department of Physics and Astronomy,\\
         Rutgers University, Piscataway, New Jersey 08854-8019, USA}
\date{\today}
\maketitle

\begin{abstract}

For the ferroelectric perovskite compounds BaTiO$_3$ and PbTiO$_3$, we have
studied the effects of external electric fields on the structural properties
of the (001) surfaces. The field-induced changes in the surface interlayer
spacings and bucklings have been calculated using a first-principles
ultrasoft-pseudopotential approach, and the change of the polarization and
the ferroelectric distortions in the surface layers have been obtained. The
surfaces are represented by periodically repeated slabs, and an external
dipole layer is included in the vacuum region of the supercells to control
the electric field normal to the surfaces. The influence of the electrical
boundary conditions on the ferroelectric properties of the slabs is discussed.
In the case of a vanishing internal electric field, our study indicates that
even very thin slabs can show a ferroelectric instability.

\end{abstract}


\vskip2pc]
\narrowtext


\section{Introduction}
\label{sec:intro}

In recent years, ferroelectric (FE) compounds based on the cubic perovskite
structure ABO$_3$ have attracted much interest because of their promising
potential for a series of technological applications. Just to mention two
examples from semiconductor industry, their ferroelectric properties can be
used to build non-volatile random access memories, and their high permittivity
makes them good candidates to replace SiO$_2$ in large-scale integrated
circuits. Because of the ongoing miniaturization in semiconductor devices,
this may be soon required in order to be able to construct capacitors with
sufficient capacity (for example in DRAM cells) and to maintain gate oxide
layers in MOSFETs which are thick enough to prevent tunneling of electrons
between the gate electrode and the channel.

In these applications it is the aim to apply the FE materials in very thin
film geometries where the ferroelectric and dielectric properties will be
strongly influenced by surface effects. For the understanding of the behavior
of FE thin films it is therefore important to explore how the FE order
parameter couples to the surface. In many experimental investigations of
thin films, a degradation of the FE properties has been observed. It has
been believed that decreasing the thickness of thin films suppresses
ferroelectricity and eliminates it altogether at a nonzero critical
thickness which has been estimated to be approximately 10~nm for
PbTiO$_3$.\cite{exp-films} However, in recent experiments, Tybell and
coworkers have shown that for high-quality films of Pb(Zr$_x$Ti$_{1-x}$)O$_3$,
a stable polarization perpendicular to the surface persists down to film
thicknesses of 40~\AA\ or less.\cite{tybell} This suggests that the
previously observed suppression of ferroelectricity in thin films is not a
purely intrinsic effect caused solely by the presence of the surface, but is
related to extrinsic factors like perturbations of the chemical composition
of the surface (impurities, oxygen vacancies or other defects), surface
induced strains, or variations in electrical and mechanical boundary
conditions.

To investigate this further, we have studied slabs of isolated, ideal
(i.e., clean and defect-free) thin films of BaTiO$_3$ and PbTiO$_3$ in (001)
orientation. Both possible surface terminations (on BaO/PbO or TiO$_2$ layers)
have been considered. It is the aim of this study to determine whether thin
films of this kind still show a ferroelectric instability, and how the
ferroelectric distortions and the spontaneous polarization change at the
surface.

A few theoretical investigations in this direction have already appeared. In
Refs.~\onlinecite{pad-ba} and \onlinecite{bm} we have studied the (001)
surfaces of BaTiO$_3$ and PbTiO$_3$ for the case of the tetragonal axis
(i.e., the direction of the spontaneous polarization) lying {\it parallel}
to the surface. Although the surface relaxation energies are found to be
substantial (i.e., many times larger than the bulk ferroelectric well depth),
it turned out that the influence of the surface upon the FE order parameter
is only modest. For the TiO$_2$--terminated surface of BaTiO$_3$ and the
PbO--terminated surface of PbTiO$_3$, a modest enhancement of the FE order
was found at the surface; for the other two surfaces, a small reduction in
the FE distortions was observed. However, in all cases the ground state was
ferroelectric, and deviations of the FE distortions from the bulk value were
confined to the first few atomic surface layers.

On the other hand, for thin PZT films the experimental investigations of
Tybell have shown a polarization {\it perpendicular} to the surface. In this
case, theoretical calculations have to deal with the additional problem of
the correct electrical boundary conditions far from the surfaces. In two
previous studies on BaTiO$_3$,\cite{cohen,resta} slabs of truncated bulk
material with a net polarization perpendicular to the surface have been used.
The slabs were either repeated periodically, assuming periodic boundary
conditions for the electrostatic potential, or were treated as isolated slabs
with a vanishing {\it external} electric field, and in neither case were
the atoms allowed to relax fully from their ideal bulk positions.

We point out in this paper that these situations are rather artificial and do
not provide useful information about the FE properties of the slabs. Instead,
we will show that the appropriate electrical boundary condition is a vanishing
electric field {\it inside} the slab. This implies that if the slab shows a
spontaneous polarization, an external electric field has to be applied to
compensate the depolarization field caused by surface charges. A vanishing
internal electric field is equivalent to the so-called {\it short-circuit}
boundary conditions, where the thin film is sandwiched between grounded plates
of a capacitor.

Short-circuit boundary conditions have been used by Ghosez and Rabe
\cite{rabe} in a microscopic effective-Hamiltonian study of PbTiO$_3$ thin
films. They found that (001)-oriented films with thicknesses as low as three
lattice constants have a perpendicularly polarized ferroelectric ground state
with a significant enhancement of the polarization at the surface. However,
in the effective Hamiltonian, no information about the structural relaxation
of the surface layers was included. With the present {\it ab initio}
calculations, we confirm the finding of Ghosez and Rabe about the
ferroelectric ground state, but we also show that the atomic relaxations at
the surface significantly modify the effective-Hamiltonian picture of the
polarization in the surface layers.

The paper is organized as follows. In Section II we describe the technical
details of our computational method and the geometry of the slabs. We also
discuss how to deduce the correct electrical boundary conditions and how to
apply these in the supercell calculations. In Section III we present the
results of our calculations on various slabs exposed to external electric
fields. Finally, the paper ends with a summary in Section IV.


\section{Theoretical Details}
\label{sec:theorie}

\subsection{Method of calculation}

As in our previous studies on perovskite surfaces,\cite{pad-ba,bm,pad-sr} we
have carried out self-consistent total-energy calculations within the
framework of density-functional theory \cite{hks} using the Vanderbilt
ultrasoft-pseudopotential scheme.\cite{van-pp}  In the pseudopotential
generation the semicore Ba 5$s$ and 5$p$, Pb 5$d$, and Ti 3$s$ and 3$p$
orbitals have been included as valence states (for more details on the
pseudopotentials see Ref.~\onlinecite{ksv}). The electron wave functions were
expanded in a plane-wave basis set including plane waves up to a cutoff
energy of 25~Ry. As has been shown in previous studies, this cutoff energy is
sufficient to obtain well converged results for these materials. A conjugate
gradient technique was used to minimize the Hohenberg-Kohn total energy
functional,\cite{ksv} and the exchange and correlation effects were treated
within the local density approximation (LDA) in the Perdew-Zunger
parameterization of the Ceperley-Alder data.\cite{ca} The positions of the
ions were determined by minimizing the atomic forces using a variable-metric
scheme.\cite{numrec} The surfaces were considered to be fully relaxed when
the forces on the ions were less than 0.01~eV/{\AA}.

\subsection{Surface and slab geometries}

BaTiO$_3$ and PbTiO$_3$ both belong to the group of II--IV perovskite
compounds, i.e., ABO$_3$ perovskites in which atoms A and B are divalent
and tetravalent, respectively. In this case, the AO and BO$_2$ layers are
charge-neutral, so that both AO-- and BO$_2$--terminated surfaces are
non-polar.

The surfaces have been represented by periodically repeated slabs consisting
of 7--9 alternately stacked layers of AO and TiO$_2$. Both types of surface
terminations have been considered, and the slabs were separated by a vacuum
region of two lattice constants. A (4,4,2) Monkhorst/Pack k-point
mesh \cite{mp} was used for all Brillouin-zone integrations. As has been
stated in Ref.~\onlinecite{bm}, the results for structural properties of
perovskite surfaces are very well converged for this choice of the supercell
geometry and k-point mesh.

BaTiO$_3$ and PbTiO$_3$ display different sequences of structural phase
transitions as the temperature is lowered. PbTiO$_3$ undergoes a single
transition from a cubic paraelectric to a tetragonal ferroelectric phase
at 763~K, which is the ground state structure at $T$=0. BaTiO$_3$ displays
a series of three transitions from cubic paraelectric to tetragonal,
orthorhombic, and rhombohedral ferroelectric phases at 403~K, 278~K, and
183~K, respectively.

Because we are primarily interested in the situation where a material shows
a polarization perpendicular to the surface, we focus here on the tetragonal
FE phases of BaTiO$_3$ and PbTiO$_3$, and we consider only the case of the
tetragonal $c$ axis pointing perpendicular to the surface (in the following
referred to as the $z$ direction). To prevent BaTiO$_3$ from adopting the true
rhombohedral $T$=0 structure, mirror symmetries M$_x$ and M$_y$ have been
imposed during the relaxation of the atoms, thus mimicking the experimental
room-temperature structure. In other words, only displacements of the atoms in
the $z$ direction (perpendicular to the surface) were allowed. This also
prevents the polarization from rotating to become parallel to the surface.
Additionally, the slab lattice constant in the $x$ and $y$ directions was set
equal to the theoretical equilibrium lattice constant $a$ computed for the
bulk tetragonal phase ($a$=3.94~\AA\ and $c$=3.99~\AA\ for BaTiO$_3$, and
$a$=3.86~\AA\ and $c$=4.04~\AA\ for PbTiO$_3$).

\subsection{Electrical boundary conditions}

For slabs with a net polarization perpendicular to the surface, two questions
have to be addressed: (i) What are the appropriate boundary conditions for
the electrostatic potential in the FE state? (ii) How will the electrostatic
potential be modified by the periodic repetition of the slabs?

\subsubsection{Boundary conditions in the FE state}
\label{sssec:FEboundary}

Let us first consider an {\it isolated} slab with a polarization perpendicular
to the surfaces. We choose the surface normal $\hat{\vc{n}}$ to be parallel to
the $z$--axis, and we assume the charge density $\rho(\vc{r})$ of the slab to
be periodic in the $x$-- and $y$--direction. The polarized slab exhibits an
electric dipole moment
\begin{equation}
\label{eq:defm}
m = \int\limits_{-\infty}^\infty \bar{\rho}(z)\,z\; {\rm d}z
\end{equation}
parallel to the surface normal $\hat{\vc{n}}$, where $\bar{\rho}(z)$ is the
planar averaged charge density
\begin{equation}
\bar{\rho}(z) = \frac{1}{A} \int\limits_{\;A}\!\!\!\int \rho(\vc{r})\;
{\rm d}x\, {\rm d}y \;\;,
\end{equation}
and $A$ is the area of the surface unit cell. The electrostatic potential
$v(\vc{r})$ seen by the electrons can be calculated by solving the Poisson
equation
\begin{equation}
\nabla^2 v(\vc{r}) = 4\pi e \rho(\vc{r}) \;\;.
\end{equation}

In addition to the microscopic quantities $\rho(\vc{r})$ and $v(\vc{r})$, we
assume the slabs to be thick enough that also macroscopic quantities like
the macroscopic electric field $\vc{E}$, the dielectric displacement field
$\vc{D}$, and the polarization $\vc{P}$ are well defined inside the slab.
In practice, these fields may be calculated, for example, from unit cell
averages of the electrostatic potential and the charge density.\cite{resta}
However, we will see that for slabs with only 7--9 atomic layers, the
applicability of this approach has its limitations.

\begin{figure}
\noindent
\epsfxsize=246pt
\epsffile{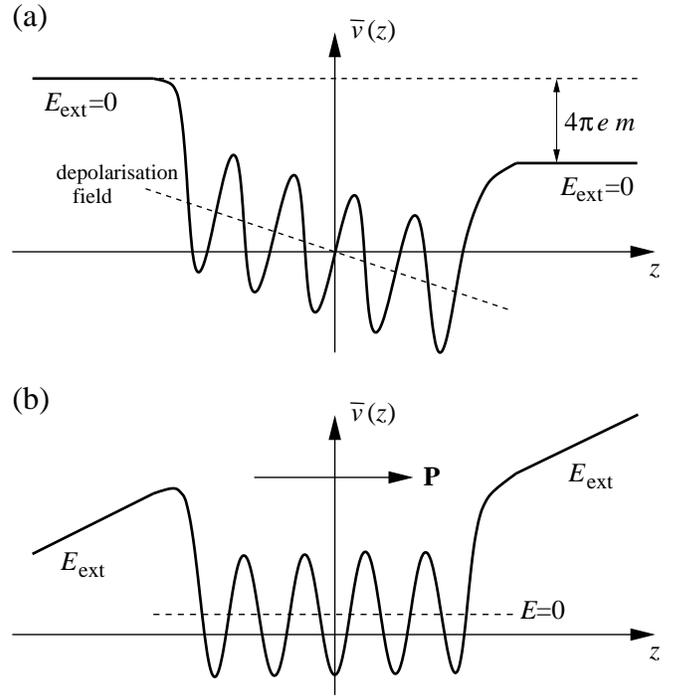}
\vspace{4pt}
\caption{\label{fig:potI}
Schematic illustration of the planar-averaged potential $\bar{v}(z)$ for an
isolated slab with a dipole moment $m$ perpendicular to the surface. (a)
Vanishing external electric field (equivalent to $D$=0). (b) Vanishing
internal electric field ($E$=0).}
\end{figure}

In the case of an applied external electric field $\vc{E}_{\rm ext}$
perpendicular to the surfaces, the dielectric displacement field $\vc{D}$
inside the slab is oriented parallel to the $z$--axis and is equal to
$\vc{E}_{\rm ext}$. The boundary condition of a vanishing external electric
field is therefore equivalent to a vanishing dielectric displacement field
$\vc{D}$ inside the slab. Figure~\ref{fig:potI}(a) shows a schematic picture
of the planar-averaged potential $\bar{v}(z)$ for this situation. The
potential is constant outside the slab, but due to the slab dipole moment $m$,
the potential jumps by $4\pi e m$ when going from one side of the slab to the
other. At the same time, the polarization $\vc{P}$ leads to surface charges
$\sigma$=$\vc{P}\!\cdot\!\hat{\vc{n}}$ which give rise to a huge
depolarization field $\vc{E}$=$\vc{D}$$-$$4\pi\vc{P}$=$-4\pi\vc{P}$ inside
the slab (notice that $\vc{E}$ does not depend on the thickness of the slab).
The contribution of the depolarization field to the total energy is large
enough to completely destabilize the bulk FE state.\cite{lines,zhong}
Therefore, relaxing a polarized slab under the boundary condition of a
vanishing external electric field will inevitably result in a paraelectric
cubic structure.

A comparison with the situation in an infinitely extended crystal shows
which boundary conditions have to be used instead. In DFT calculations for
bulk systems, periodic boundary conditions for the electrostatic potential
are usually applied. In this case, the internal electric field $\vc{E}$
vanishes, even in the presence of a spontaneous polarization
$\vc{P}_{\rm s}^{\rm bulk}$, whereas the dielectric displacement field
$\vc{D}$ will be nonzero. In analogy, a slab is in a FE state with spontaneous
polarization $\vc{P}_{\rm s}$ if a situation exists where the internal
electric field $\vc{E}$ is zero, but the dielectric displacement field
$\vc{D}$ is nonzero. However, an external electric field
$\vc{E}_{\rm ext}$=$\vc{D}$=$4\pi\vc{P}_{\rm s}$ will then appear outside the
slab, as shown in Fig.~\ref{fig:potI}(b).

In our calculations we are only able to control the external electric field
$\vc{E}_{\rm ext}$ but not the internal field $\vc{E}$. So to study whether a
slab shows a FE instability, we have to apply external electric fields of
different strength and search for the situation where
$\vc{E}$=$\vc{E}_{\rm ext}$$-$$4\pi\vc{P}$ is zero. A rough estimate of how
large $\vc{E}_{\rm ext}$ will be in this situation can be made if we assume
that the polarization of the slabs is equal to the bulk spontaneous
polarization $\vc{P}_{\rm s}^{\rm bulk}$. By using the Berry phase
approach,\cite{berry} we have calculated the bulk spontaneous polarization of
BaTiO$_3$ and PbTiO$_3$ to be 4.6$\times$10$^{-3}$~e/bohr$^2$ and
$14.2\times$10$^{-3}$~e/bohr$^2$, respectively. This translates to external
electric fields of 0.058~a.u.\ and 0.18~a.u.\ (atomic units are used
throughout the paper\cite{units}).

\begin{figure}
\noindent
\epsfxsize=246pt
\epsffile{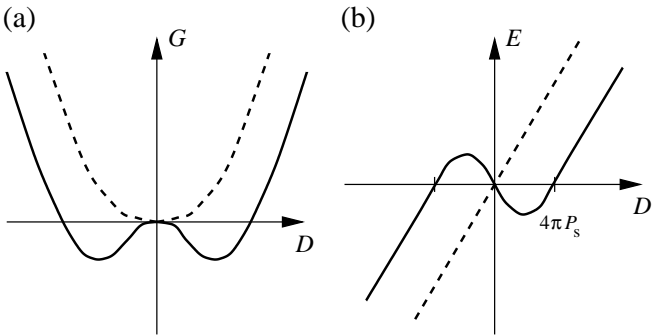}
\vspace{4pt}
\caption{\label{fig:thermo}
Sketch of the Gibbs free energy $G$ and the internal electric field $E$ as a
function of the dielectric displacement $D$ (or the external electric field
$E_{\rm ext}$, respectively) for a paraelectric (dotted line) or ferroelectric
(solid line) slab.}
\end{figure}

To be able to distinguish between a paraelectric or ferroelectric behavior of
a slab, we employ a simple phenomenological picture. We introduce the elastic
Gibbs energy\cite{lines} $G(\vc{D})$ and we concentrate only on the dependence
on the dielectric displacement field $\vc{D}$ (neglecting temperature and
strain effects this is just the internal energy $U$). For a paraelectric,
$G(\vc{D})$ is roughly quadratic in $\vc{D}$, whereas for a FE material
$G(\vc{D})$ will show the well known double-well structure (see
Fig.~\ref{fig:thermo}(a)). Differentiating the Gibbs function gives
immediately the internal electric field
\begin{equation}
\vc{E} = \frac{\partial G}{\partial \vc{D}} \;\;.
\end{equation}
Thus, calculating $\vc{E}(\vc{D})$ directly reveals whether a slab is FE or
not (see Fig.~\ref{fig:thermo}(b)). For a paraelectric slab, $\vc{E}$ is
proportional to $\vc{D}$, with the proportionality factor given by the
reciprocal dielectric constant, for a ferroelectric, the internal electric
field is first negative as the polarization builds up, and then vanishes when
the spontaneous polarization $\vc{P}_{\rm s}$ is reached. We will make use of
this kind of analysis later in Secs.~\ref{sec:norelax} and \ref{sec:ferinst}.

\subsubsection{Periodically repeated supercells}

\begin{figure}
\noindent
\epsfxsize=246pt
\epsffile{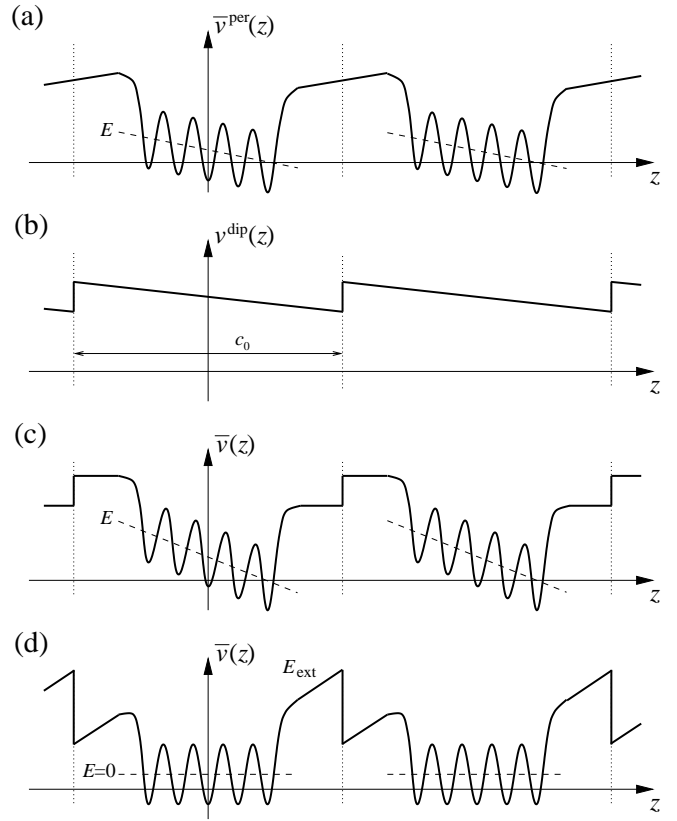}
\vspace{4pt}
\caption{\label{fig:potII}
Schematic picture of the planar averaged potential $\bar{v}(z)$ for
periodically repeated slabs.
(a) with periodic boundary conditions,
(b) potential of the dipole layer,
(c) dipole corrected slabs with vanishing external electric field,
(d) dipole corrected slabs with vanishing internal electric field.}
\end{figure}

In supercell calculations, periodic boundary conditions are usually imposed
on the electrostatic potential. For slabs with a non-vanishing dipole moment
perpendicular to the surface, this leads to electrostatic potentials that
typically look like the sketch shown in Fig.~\ref{fig:potII}(a). The
electrostatic potential corresponds neither to the situation of
Fig.~\ref{fig:potI}(a) nor Fig.~\ref{fig:potI}(b). Instead, the imposition
of periodic boundary conditions on the supercell geometry leads to some other
particular combination of internal and external electric fields, such that
there is no discontinuity in the potential at the supercell boundary. The same
occurs even for paraelectric slabs when terminated by nonequivalent surfaces
with different work functions.

The artificial electric fields become smaller when the thickness of the slab
or the vacuum region is increased, but it is computationally very expensive
to converge results by using larger and larger supercells. Fortunately, the
error associated with the artificial electric field can easily be eliminated
by introducing an external dipole layer in the vacuum region of the
supercell.\cite{bengt,neug} The electrostatic potential of this dipole layer
is shown in Fig.~\ref{fig:potII}(b). In order to reach a situation
corresponding to Fig.~\ref{fig:potI}(a), the unwanted artificial external
electric field can be compensated by {\it adding} a certain amount of the
potential of Fig.~\ref{fig:potII}(b) to that of Fig.~\ref{fig:potII}(a), as
is shown in Fig.~\ref{fig:potII}(c) (dipole correction\cite{bengt}). The
dipole-corrected electrostatic potential is now discontinuous, but the
discontinuity lies in the vacuum region of the supercell where the wave
functions are essentially zero. Alternatively, the external dipole layer may
be used to apply a true external electric field $\vc{E}_{\rm ext}$ to the
surfaces.\cite{neug} In particular, the situation of Fig.~\ref{fig:potI}(b)
can be reached by {\it subtracting} a certain amount of the dipole potential
of Fig.~\ref{fig:potII}(b) from that of Fig.~\ref{fig:potII}(a), as shown
in Fig.~\ref{fig:potII}(d).

The external dipole field can easily be implemented in any plane-wave based
electronic structure code. Following the notation of Bengtsson,\cite{bengt}
we denote the external dipole potential of Fig.~\ref{fig:potII}(b) as
$v^{\rm dip}(z)$, and the electrostatic potential for the electrons calculated
under periodic boundary conditions (corresponding to Fig.~\ref{fig:potII}(a))
as $v^{\rm per}(\vc{r})$. The new potential is then
\begin{equation}
v(\vc{r}) = v^{\rm per}(\vc{r}) + v^{\rm dip}(z) \;\;.
\end{equation}
For a slab with dipole moment $m$ and an external electric field
$E_{\rm ext}$, the dipole potential is given by
\begin{equation}
v^{\rm dip}(z) = -e \left( \frac{4\pi m}{c_0} - E_{\rm ext} \right) z \;\;,
\quad -\frac{c_0}{2}<z<\frac{c_0}{2} \;\;,
\end{equation}
where $c_0$ is the height of the supercell. In a self-consistent calculation,
the charge density, and thereby $m$, change with each step of the iteration.
Therefore, $m$ and the dipole potential $v^{\rm dip}(z)$ have to be
recalculated on each iteration until self-consistency is achieved (analogous
to the updating of the Hartree and the exchange-correlation potentials).

The additional external potential $v^{\rm dip}(z)$ also leads to changes in
the total energies $E_{\rm tot}$ and the Hellmann-Feynman forces $\vc{F}_I$:
\begin{equation}
E_{\rm tot} = E_{\rm tot}^{\rm per} + \left( \frac{2\pi m}{c_0} - E_{\rm ext}
\right) A\, m\;\;,
\end{equation}
\begin{equation}
\vc{F}_I = \vc{F}_I^{\rm per} - e Z_I \left( \frac{4\pi m}{c_0} - E_{\rm ext}
\right) \hat{\vc{e}}_z \;\;,
\end{equation}
where $E_{\rm tot}^{\rm per}$ and $\vc{F}_I^{\rm per}$ are the total energy
and Hellmann-Feynman force calculated with the periodic potential
$v^{\rm per}(\vc{r})$, and $Z_I$ is the ionic charge of ion $I$.


\section{Results and Discussion}
\label{sec:results}

\subsection{Zero external electric field}
\label{sec:nofield}

As a first step, we calculated the fully relaxed structure of the various
slabs in zero external electric field. For the asymmetrically terminated
8-layer slabs, the dipole correction was used to enforce the vanishing
of the external field. In this case, as has been pointed out in
Sec.~\ref{sssec:FEboundary}, all slabs adopt the paraelectric cubic phase.
The symmetrically terminated slabs show no dipole moment (the central layer
of the slabs is a mirror plane), and the small dipole moment of the
asymmetrically terminated slabs is solely caused by the difference between
the work functions of the two surfaces.

Detailed results on the structure of the BaTiO$_3$ and PbTiO$_3$ surfaces
in the cubic phase in the absence of an external electric field have already
been published in Refs.~\onlinecite{pad-ba} and \onlinecite{bm}. Therefore,
we give here only a brief summary in order to establish notation and to
provide a baseline for comparison.

\begin{figure}
\noindent
\epsfxsize=246pt
\epsffile{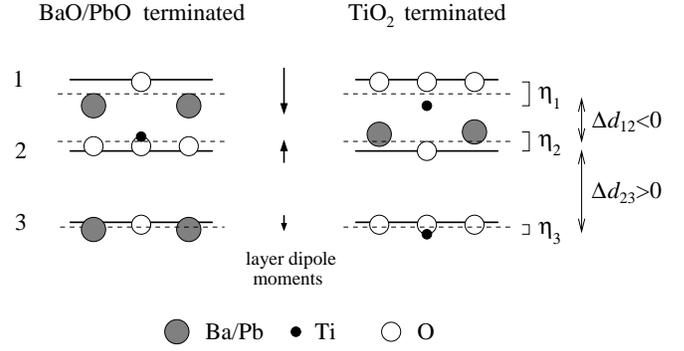}
\vspace{4pt}
\caption{\label{fig:struc}
Schematic illustration of the structure of the first three surface layers.}
\end{figure}

Figure~\ref{fig:struc} shows a schematic illustration of the structure of the
first three surface layers, where the relaxations of the atoms have been
highly exaggerated. For both BaTiO$_3$ and PbTiO$_3$, and for both the BaO/PbO
and TiO$_2$ surface terminations, the same characteristic features appear
regarding the buckling of the surface layers and the changes of the interlayer
distances. Only the amplitudes of these relaxations differ from surface to
surface. For a quantitative analysis of the surface relaxations, we let
$\delta_z$(O$_i$) and $\delta_z$(M$_i$) be the displacements of the oxygen
and metal atoms, respectively, relative to the ideal unrelaxed structure
in layer $i$. The change of the interlayer distance $\Delta d_{ij}$ is then
given by the difference of the averaged displacements
($\delta_z$(M)+$\delta_z$(O))/2 of the atoms in layers $i$ and $j$. For all
surfaces, $\Delta d_{12}$ is negative (corresponding to a reduction of the
interlayer distance between the first and the second surface layer compared
to the bulk value), whereas $\Delta d_{23}$ is positive. To describe the
buckling of the surface layers, we define a rumpling parameter $\eta_i$ as
the amplitude of the relative displacements between the metal and the oxygen
ions: ($\delta_z$(M$_i$)$-$$\delta_z$(O$_i$)). $\eta_i$ is negative
if the metal ions are below the oxygen atoms, which is true for the first
surface layer of all surfaces. For the next surface layers, $\eta_i$
oscillates in sign from layer to layer, and the amplitude decreases very
rapidly.

\subsection{External electric field without field-induced atomic relaxations}
\label{sec:norelax}

In the next step, we exposed the slabs to external electric fields
$\vc{E}_{\rm ext}$ of increasing strength, but we kept all atoms frozen in
the positions which have been calculated in zero external field. Only the
distribution of the electrons was recalculated self-consistently. This mimics
the exposure of the slabs to a AC electric field with a frequency high enough
that the ions are no longer able to follow. In this situation, we expect the
slabs to show a purely paraelectric response.  That is, a linear relation
between the internal electric field $\vc{E}$ and the dielectric displacement
field $\vc{D}$ should be found, $\vc{D}$=$\epsilon_\infty$$\vc{E}$, where
$\epsilon_\infty$ is the optical dielectric constant.

\begin{figure}
\noindent
\epsfxsize=246pt
\epsffile{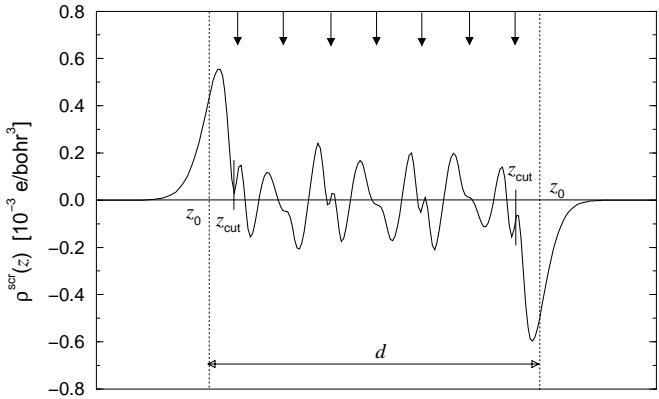}
\vspace{4pt}
\caption{\label{fig:screencharge}
Planar-averaged screening charge distribution of a BaO--terminated slab
of BaTiO$_3$, calculated from a 7-layer slab for an electric field of
$E_{\rm ext}$=+0.02~a.u. The arrows indicate the positions of the atomic
planes. ``Top'' of slab is at right.}
\end{figure}

To verify this, we calculated the internal electric field $\vc{E}$ as a
function of the applied external electric field $\vc{E}_{\rm ext}$ (or,
equivalently, the dielectric displacement field $\vc{D}$). A natural way to
compute $\vc{E}$ would be to determine the gradient of the macroscopically
averaged electrostatic potential $v(\vc{r})$. Unfortunaly, it turned out that
7--9 layers are not enough to give accurate macroscopic averages (the
uncertainties in the gradients were much too large). Instead, we calculated
the internal electric field by using the relation
$\vc{E}$=$\vc{D}$$-$$4\pi\vc{P}$. The polarization $\vc{P}$ can either be
deduced from the surface charge $\sigma$ or from the dipole moment $m$ and the
thickness $d$ of the slabs:
\begin{equation}
\label{eq:defp}
P = \sigma \;\;, \qquad P = m/d \;\;.
\end{equation}
The dipole moment $m$ is directly given by the charge distribution of the
slabs via Eq.~(\ref{eq:defm}). For the thickness $d$ we have taken the
distance between the centers of gravity $z_0$ of the screening charge
distribution of the top and bottom surface of our slabs. (For a metal, $z_0$
is the position of the surface image plane from which the classical image
potential is measured.) The screening charge distribution is given by the
difference of the charge densities calculated with and without an external
electric field. A typical planar-averaged screening charge distribution
$\bar{\rho}^{\,\rm scr}(z)$ is shown in Fig.~\ref{fig:screencharge}. As
expected, a positively-oriented electric field pushes the electrons into the
slab at the top surface and pulls them out at the bottom (where it can be
interpreted as a negatively-oriented field). From Fig.~\ref{fig:screencharge},
we see that the screening charge piles up directly at the surface and is
mainly confined above the surface atoms. These surface charges are responsible
for the screening of the external electric field. Inside the slab, fairly
large oscillations are found which are associated with the remaining internal
electric field.

\begin{figure}
\noindent
\epsfxsize=246pt
\epsffile{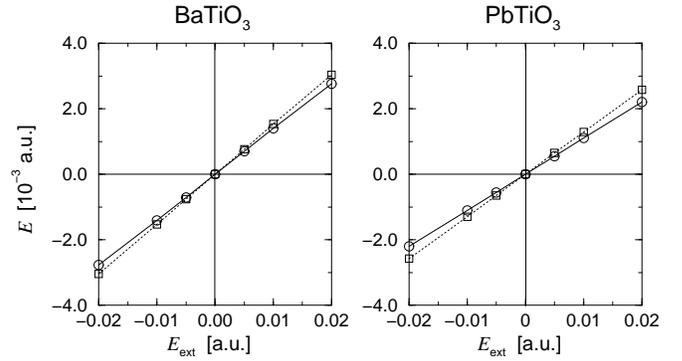}
\vspace{4pt}
\caption{\label{fig:eint_unrelax}
Internal electric field $E$ in the 7-layer slab with frozen atomic positions
as a function of the applied external electric field $E_{\rm ext}$. Solid
lines: AO--terminated slabs. Dotted lines: TiO$_2$--terminated slabs. Atomic
units are used.\cite{units}}
\end{figure}

\begin{table}
\noindent
\begin{tabular}{ccccc}
     & \multicolumn{2}{c}{BaTiO$_3$} & \multicolumn{2}{c}{PbTiO$_3$}\\
     & BaO--term. & TiO$_2$--term. & PbO--term. & TiO$_2$--term.\\ \hline
  7  &        7.1  &        6.5  &        9.0  &        7.7 \\
  8  &             &             &  \multicolumn{2}{c}{8.1} \\
  9  &        7.0  &        6.6  &             &            \\
bulk &  \multicolumn{2}{c}{6.75} &  \multicolumn{2}{c}{8.24}
\end{tabular}
\caption{\label{tab:eps}
Calculated optical dielectric constants $\epsilon_\infty$ for the various
slabs. The theoretical bulk values are taken from
Ref.~\protect\onlinecite{rabedat}.}
\end{table}

The surface charge $\sigma$ and the center of gravity of the screening charge
distribution $z_0$ for the top surface of the slab are given by
\begin{equation}
\label{def_d}
\sigma = \int\limits_{z_{\rm cut}}^{c_0/2} \bar{\rho}^{\,\rm scr}(z)\;
{\rm d}z \;\;,\qquad z_0 = \frac{1}{\sigma} \int\limits_{z_{\rm cut}}^{c_0/2}
\bar{\rho}^{\,\rm scr}(z)\,z\; {\rm d}z \;\;.
\end{equation}
For $z_{\rm cut}$ we have taken the position where the extrapolation of the
first charge peak goes to zero (see Fig.~\ref{fig:screencharge}), but the
results are not very sensitive to the choice of $z_{\rm cut}$. Analogous
relations apply for the bottom surface. To check the consistency of this
approach, we have calculated the polarization $\vc{P}$ of the slabs in both
of the ways indicated in Eq.~(\ref{eq:defp}), i.e., using either the surface
charges $\sigma$ or the dipole moment $m$ with the thickness $d$. In all
cases, the computed values for the polarization differ by less than 1\%.

\widetext
\begin{figure*}
\noindent
\epsfxsize=246pt
\epsffile{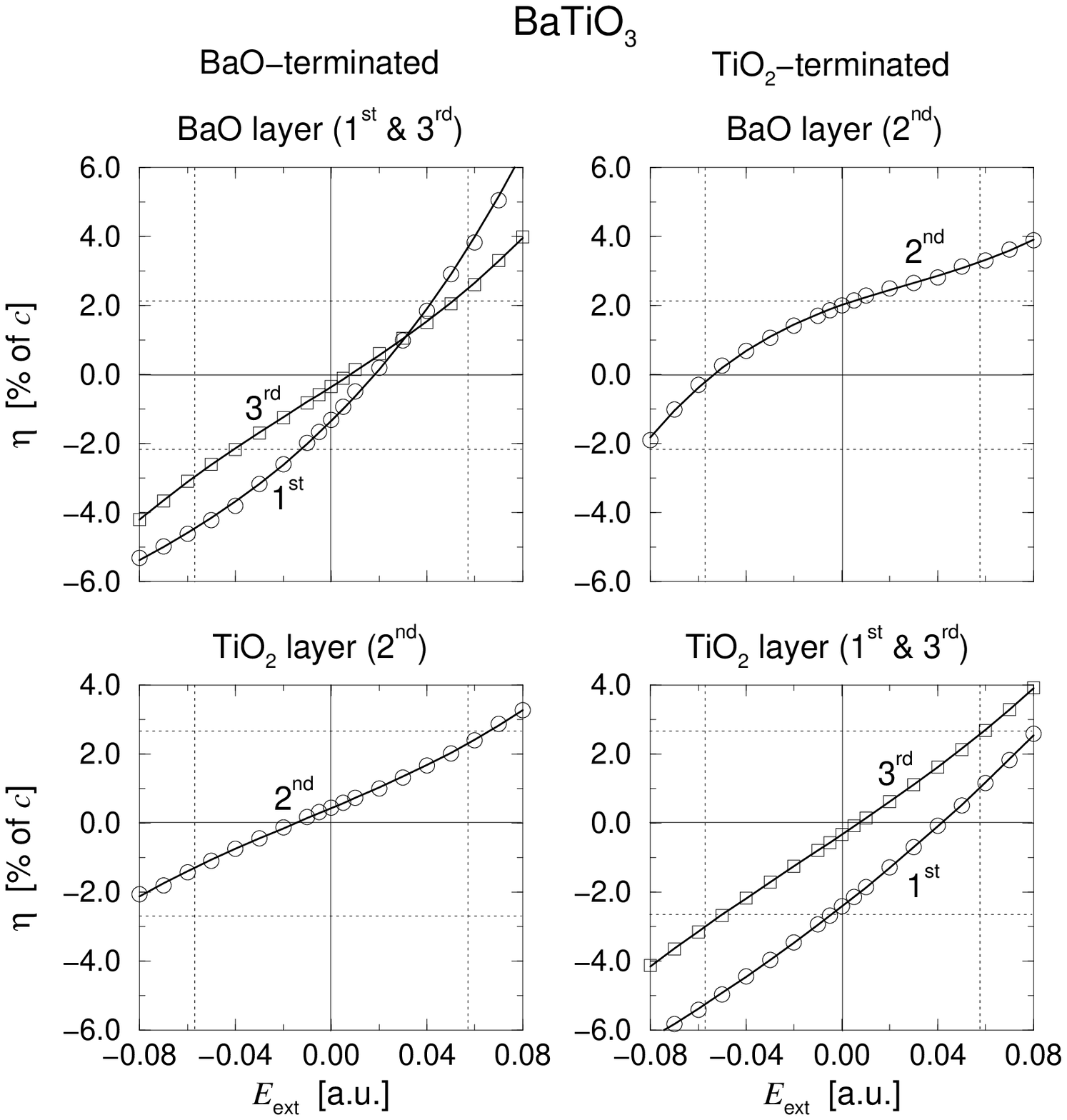}
\hfill
\epsfxsize=246pt
\epsffile{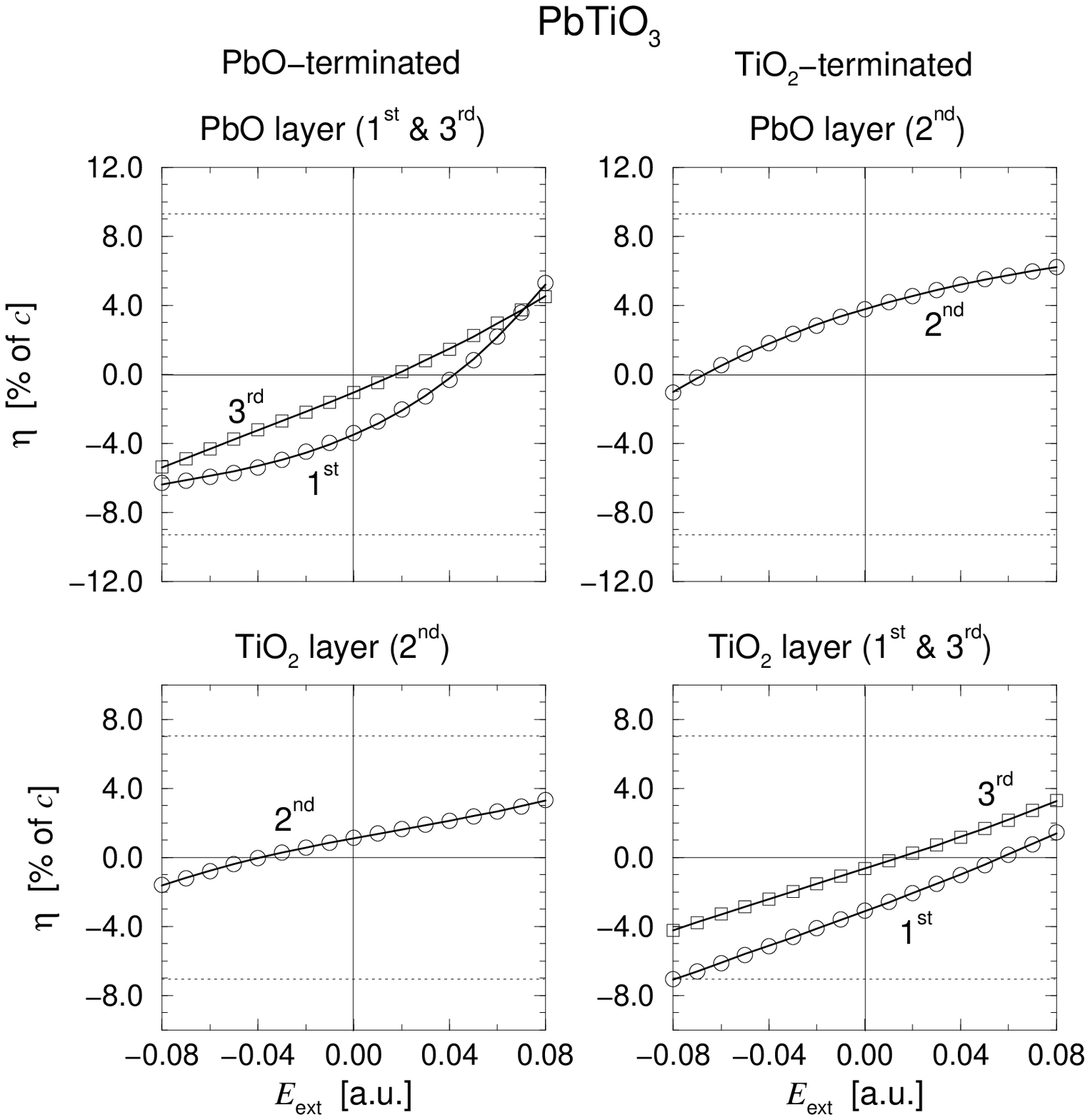}
\vspace{4pt}
\caption{\label{fig:eta}
Field dependence of the rumpling parameter $\eta_i$ calculated from the
7-layer slab. Dotted horizontal lines: amplitude of displacement of metal and
oxygen ions in the bulk ferroelectric state. Dotted vertical lines: External
electric field corresponding to the bulk spontaneous polarization.}
\end{figure*}
\narrowtext

With the polarization $\vc{P}$ we can finally determine the internal electric
field $\vc{E}$. Figure~\ref{fig:eint_unrelax} shows the internal electric
field for the 7-layer slabs as a function of the applied external field.
Compared with our phenomenological picture in Fig.~\ref{fig:thermo}, a clear
paraelectric behavior can be seen for all slabs. Table~\ref{tab:eps} gives the
optical dielectric constants deduced from the $E(D)$ plots for the different
slabs, together with the LDA bulk values of Ref.~\onlinecite{rabedat}. Already
for 7 atomic layers, the dielectric constants agree very well with the bulk
values. The finite-size effect is less than 10\%. Its sign depends on the
relative numbers of AO and TiO$_2$ layers: $\epsilon_\infty$ is increased when
the number of AO--layers exceeds the number of TiO$_2$--layers, and is reduced
when the slabs contain more TiO$_2$ layers.

\subsection{External electric field with atomic relaxations}

Finally, all slabs were fully relaxed in the presence of external static
electric fields. Fields of strength varying between $-$0.08~a.u.\ and
$+$0.08~a.u.\ have been applied. Starting from zero field, the amplitude of
the field was changed in steps of 0.01~a.u., and the atomic positions at the
previous field were used as initial configurations for the new atomic
relaxations. The upper limit of 0.08~a.u.\ for the amplitude of the external
electric field is determined in our calculations by the thickness of the
vacuum region for the following reason.  As can be seen in
Fig.~\ref{fig:potII}(d), there is a kind of ``quantum well'' located in
the vacuum region just to the right of the dipole layer. If the electrostatic
potential of this quantum well drops below the Fermi level, it can become
populated by the transfer of electrons from the slab region. The threshold
for the occurrence of this unwanted behavior depends on both the width of the
vacuum region and the strength of the electric field.

\subsubsection{Field-induced atomic relaxations}

In a positively oriented external electric field, the negatively charged
oxygen ions will be pushed into the bulk and the positive ions will be pulled
towards the surface. This will immediately change the rumpling parameter
$\eta_i$ of the surface layers, as can be seen in Fig.~\ref{fig:eta}. For
small applied fields, $\eta_i$ increases linearly with the strength of
$E_{\rm ext}$, as is expected if a harmonic coupling between the atoms is
assumed. However, for larger fields, nonlinear effects become important,
especially for the first and second AO layers.

To facilitate comparison of these surface rumplings with the corresponding
bulk rumplings of the ``up'' and ``down'' FE states, the latter are drawn as
horizontal dotted lines in Fig.~\ref{fig:eta}. That is, these lines show the
values that the rumpling parameters would have if the corresponding slab had
been a piece of truncated FE bulk material without any further relaxation of
the atoms. Additionally, for BaTiO$_3$ we have used dotted vertical lines to
indicate the magnitude of the external electric field
${\bf E}_{\rm ext}$=$4\pi{\bf P}_{\rm s}^{\rm bulk}$ that would be consistent
with a bulk spontaneous polarization $\vc{P}_{\rm s}^{\rm bulk}$ in the
interior of the slab.

It can be seen from Fig.~\ref{fig:eta} that for this external electric field,
the rumpling parameters $\eta_i$ of the surface layers have reached the same
order of magnitude as the bulk FE distortions. For most layers $\eta_i$ is
slightly larger, but for some layers it is a little bit smaller. Overall, the
rumpling parameter $\eta_i$ tends to be larger for the negative electric
field than for the positive field, suggesting that the surface polarization
is larger for the negatively-oriented field. However, it is difficult to
deduce from Fig.~\ref{fig:eta} whether the polarization is enhanced or reduced
at the surface compared with the bulk. We will return to this question in the
next subsection.

The bulk spontaneous polarization $\vc{P}_{\rm s}^{\rm bulk}$ is roughly
three times larger for PbTiO$_3$ than for BaTiO$_3$
(see Section~\ref{sssec:FEboundary}). The external electric field for the
case that a slab adopts the PbTiO$_3$ bulk spontaneous polarization is
therefore on the order of 0.18~a.u. This is well beyond the upper limit of
the electric field that can be applied in our calculations. Consequently, the
polarization of the PbTiO$_3$ slabs is well below $\vc{P}_{\rm s}^{\rm bulk}$
for all applied external electric fields, and therefore it is no surprise that
the surface rumpling $\eta_i$ in Fig.~\ref{fig:eta} is much smaller than the
corresponding bulk FE rumpling over the entire range of accessible external
electric fields.

\begin{figure}
\noindent
\epsfxsize=246pt
\epsffile{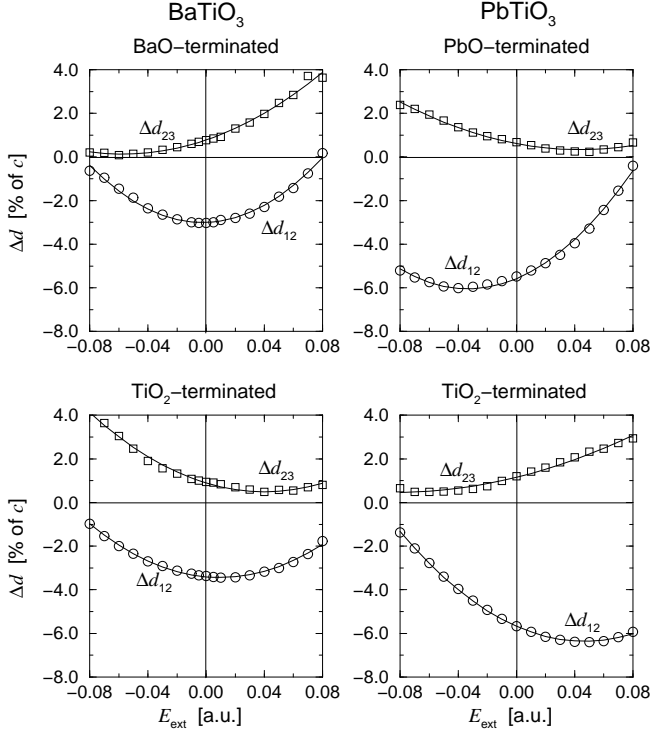}
\vspace{4pt}
\caption{\label{fig:delta_d}
Field-dependence of the change in interlayer distances $\Delta d_{ij}$
calculated for the 7-layer slabs.}
\end{figure}

Associated with the changes in the layer rumplings, we also observe changes in
the interlayer distances as shown in Fig.~\ref{fig:delta_d}. (Note that the
atomic layers are charge-neutral, so that there is no net force on the layers
in an external electric field.) We find some noteworthy qualitative
differences in the behaviors of the two materials (e.g., in the slopes of the
$\Delta d_{23}$ {\it vs.}~$E_{\rm ext}$ curves). Ultimately these differences
can presumably be traced to the rather different chemistry of Ba--O and Pb--O
bonds, but a detailed explanation of the observed trends is not obvious.

\subsubsection{Field-induced layer dipoles}

To address the question how the polarization is changed at the surface, we
have subdivided the total dipole moment $m$ of the slabs from
Eq.~(\ref{eq:defm}) into dipole moments $m_i$ per atomic layer. This
was done by identifying the ``nodes'' $z_i$ at which
$\int_{-c_0/2}^{z_i} \bar{\rho}(z)\,{\rm d}z$ vanishes. Regions between
successive nodes are thus charge neutral and contain exactly one AO or
TiO$_2$ atomic plane. The dipole moment $m_i$ was then calculated via
Eq.~(\ref{eq:defm}) while restricting the range of integration to the area
between the two nodes which embrace the atomic layer of interest.

\begin{figure}
\noindent
\epsfxsize=246pt
\epsffile{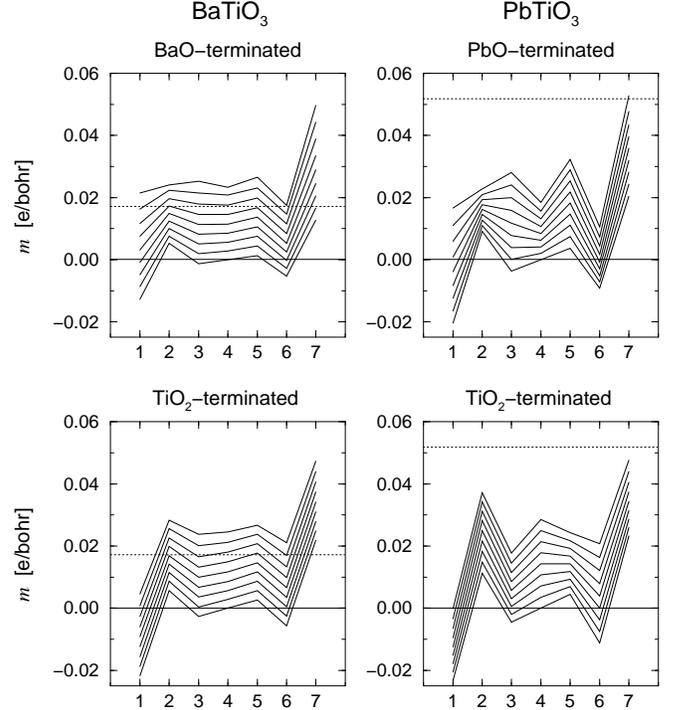}
\vspace{4pt}
\caption{\label{fig:dipole}
Dipole moment per atomic layer calculated for the 7-layer slabs for electric
fields from 0 to 0.08~a.u.\ in steps of 0.01~a.u. The orientation of the
electric field corresponds to a positive field for the top surface and a
negative field for the bottom surface. The dotted line represents the dipole 
moment for an atomic layer in the bulk ferroelectric state, calculated from
the bulk spontaneous polarization via $c$/2$\cdot P_{\rm s}^{\rm bulk}$.}
\end{figure}

Fig.~\ref{fig:dipole} shows the dipole moments per atomic layer for the
7-layer slab in different external electric fields. For zero external electric
field, the dipole moments oscillate in sign from layer to layer. The dipole
moment is largest for the first atomic layer at the surface and goes to zero
very rapidly (see also Fig.~\ref{fig:struc}). Finally, the net dipole moment
of the surface, given by the sum of the $m_i$ of all surface layers, points
inwards towards the bulk for all surfaces.

An external electric field causes the dipole moments of all layers in the
slab to increase more or less equally. The dipole moments continue to
oscillate from layer to layer when $E_{\rm ext}$ is increased, but they
approach a constant bulk value already after two layers in BaTiO$_3$ and after
three layers in PbTiO$_3$. (This becomes more apparent from the 8- and 9-layer
slab calculations, not shown here.) The dipole moments in the surface layers
are modified relative to the bulk layers by the inwards oriented surface
dipoles. This leads to a suppression of the polarization at the surface in a
positively-oriented external electric field, and an enhancement for a negative
field. However, the modification of the polarization is more or less confined
to the first unit cell (i.e., first pair of layers) at the surface.

\begin{table}
\noindent
\begin{tabular}{ccccccc}
          & \multicolumn{2}{c}{dipole moment} &
            \multicolumn{4}{c}{polarization} \\
          & AO--term. & TiO$_2$--term. & \multicolumn{2}{c}{AO--term.} &
            \multicolumn{2}{c}{TiO$_2$--term.}\\ \hline
SrTiO$_3$ &   $-$8.7  &     $-$14.2    & 1.19 & (--)    & 1.95 & (--) \\
BaTiO$_3$ &   $-$7.4  &     $-$14.7    & 0.98 & (21\%)  & 1.95 & (42\%) \\
PbTiO$_3$ &  $-$11.3  &     $-$11.8    & 1.48 & (10\%)  & 1.55 & (11\%)
\end{tabular}
\caption{\label{tab:dip}
Dipole moment (in 10$^{-3}$e/bohr) and polarization (in 10$^{-3}$e/bohr$^2$)
of the surface unit cell as calculated from the 7-layer slabs.\cite{units}
The surface polarization as a fraction of the theoretical bulk spontaneous
polarization is given in parentheses.  Values for SrTiO$_3$ are given for
comparison.}
\end{table}

In most cases a rough estimate of how much the polarization is suppressed or
enhanced at the surface is given by the surface dipole moments in zero
external electric field. Table~\ref{tab:dip} lists the dipole moment of the
uppermost unit cell (i.e., the sum of the dipole moments $m_i$ of the first
two atomic layers) in zero external electric field for slabs of 7 layers.
(The results from 8- and 9-layer slabs differ by less than 3\%.) Additionally,
these dipole moments have been converted to a polarization by dividing them by
the lattice constant $c$.

The values of the dipole moments of the uppermost unit cells are of the same
order of magnitude for all surfaces. In the case of BaTiO$_3$, they correspond
to about 20--40\% of the bulk spontaneous polarization
$\vc{P}_{\rm s}^{\rm bulk}$. Therefore, for BaTiO$_3$ a relatively large
change in the polarization at the surface has to be expected for a
spontaneously polarized slab. On the other hand, for PbTiO$_3$ the surface
dipoles make up only about 10\% of the bulk spontaneous polarization. The
modification of the polarization at the surface in a spontaneously polarized
state will therefore be only modest. The polarization will be suppressed for
an upwards polarization and enhanced for a downwards polarized state.

\subsubsection{Ferroelectric instability of the slabs}
\label{sec:ferinst}

The results presented so far on the structure of the surfaces do not provide
an answer to the question whether the slabs exhibit a true FE instability
or simply show a paraelectric polarization. Instead, we have to determine the
electric field inside the slabs as a function of the applied external field
and compare the result with our simple phenomenological picture from
Fig.~\ref{fig:thermo}(b).

\begin{figure}
\noindent
\epsfxsize=246pt
\epsffile{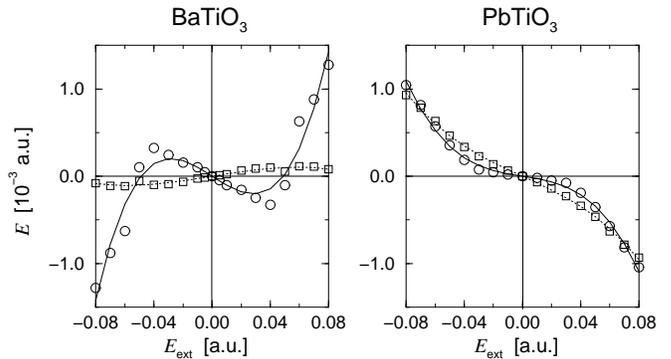}
\vspace{4pt}
\caption{\label{fig:eint_relax}
Internal electric field $E$ in the fully relaxed 7-layer slab as a function
of the applied external electric field $E_{\rm ext}$. Solid lines,
AO--terminated slabs; dotted lines, TiO$_2$--terminated slabs.}
\end{figure}

For the calculation of the internal electric field we followed the same
procedure as described in Section~\ref{sec:norelax}. However, instead of using
the true screening charge distribution (which includes contributions from the
moving ions) to determine $z_0$ and $d$, we took the difference between the
charge density of the fully relaxed slab in the external electric field and
the charge density of a calculation where the atomic positions have been kept
fixed and the electric field turned off. Unfortunately it turned out that it
is not possible to determine the internal electric field to very high
precision. The use of the relation $\vc{E}$=$\vc{D}$$-$$4\pi\vc{P}$ involves
the difference of two relatively large numbers, so that the result for the
internal electric field depends very sensitive on details of the calculation.
In particular, the determination of the thickness $d$ of the slabs is very
critical. Also when approaching the breakdown field above which electrons
accumulate in the quantum well next to the dipole layer, the wavefunctions
start to penetrate the barrier between the slab and the quantum well. This
leads to a small overestimation of the dipole moment and the polarization
of the slab at high applied fields. But because of the subtle difference
between the external electric field and the polarization, the error in the
internal electric field becomes noticable, and deviations from the behavior
described in Fig.~\ref{fig:thermo}(b) may appear at large external fields.
Therefore, the curves of the internal electric field as a function of the
applied external electric field shown in Fig.~\ref{fig:eint_relax} represent
only basic tendencies but are not to be taken as quantitatively accurate
results.

The results in Fig.~\ref{fig:eint_relax} are clearest for the BaO--terminated
slab of BaTiO$_3$, where the full curve expected for ferroelectric behavior,
as in Fig.~\ref{fig:thermo}(b), can be observed. We find that the internal
electric field vanishes for an external electric field of approximately
0.05~a.u. (the result for the 9-layer slab is the same). This translates to
a spontaneous polarization of the slab of
$P_{\rm s}$=4$\times$10$^{-3}$~e/bohr$^2$, which corresponds roughly with
the bulk spontaneous polarization $\vc{P}_{\rm s}^{\rm bulk}$ of BaTiO$_3$.
However, due to the limitations in the accuracy of our calculations, we are
not able to say for certain whether the spontaneous polarization of the slab
is enhanced or reduced compared to the bulk value.

For the other three cases in Fig.~\ref{fig:eint_relax} it is more difficult
to deduce, whether the slabs show a FE instability or not. For PbTiO$_3$ we
have the problem that the upper limit for the external electric field is
reached long before the slabs come close to a polarization equal to the bulk
spontaneous polarization (as has been pointed out above). In
Fig.~\ref{fig:eint_relax}, the external electric field where the internal
field is zero can therefore not be reached for the PbTiO$_3$ slabs.

On the other hand, according to Fig.~\ref{fig:thermo}(b), a negative slope of
the internal electric field at zero applied field is already a very good
indicator for a FE instability. As can be seen from Fig.~\ref{fig:eint_relax},
this condition is fulfilled for both the PbO--terminated and the
TiO$_2$--terminated slab of PbTiO$_3$, but not for the TiO$_2$--terminated
slab of BaTiO$_3$. To estimate how sensitive this result depends on details
of our calculations we compare the numerical values of the initial slopes of
the $E(D)$ curves with an error bar which we derive from the uncertainties in
the determination of the slab thickness $d$. The fit curves in
Fig.~\ref{fig:eint_relax} yield initial slops of $-$9.1$\times$10$^{-3}$ and
+2.6$\times$10$^{-3}$ for the BaO-- and TiO$_2$--terminated BaTiO$_3$ slabs
and $-$2.5$\times$10$^{-3}$ and $-$7.6$\times$10$^{-3}$ for the PbO-- and
TiO$_2$--terminated PbTiO$_3$ slabs, respectively. By varying $z_{\rm cut}$
in Eq.~(\ref{def_d}) we can estimate an error bar of about
$\pm$4$\times$10$^{-3}$. For the BaO--terminated BaTiO$_3$ slab and the
TiO$_2$--terminated PbTiO$_3$ slab the initial slopes are well outside the
error bar, indicating a ferroelectric instability of the slabs, whereas for
the TiO$_2$--terminated BaTiO$_3$ slab and the PbO--terminated PbTiO$_3$ slab
our analysis of the internal electric field is not accurate enough to make a
clear statement. It would be possible to clarify the situation in the last
two cases by going to thicker slabs with more atomic layers, since then the
thickness $d$ would be relatively less uncertain, but this was deemed too
computationally demanding to be undertaken in the current project. At least
we can say that both slabs are very close to a ferroelectric instability.


\section{Summary}
\label{summary}

By using first-principles density-functional calculations, we have studied
(001)-oriented slabs of BaTiO$_3$ and PbTiO$_3$ in external electric fields.
Artificial electric fields introduced by the periodic boundary conditions of
the supercell approach have been eliminated with the help of an external
dipole layer in the vacuum region. In this way it is possible to handle
periodically repeated polarized slabs as if they were genuinely isolated,
without the need for large vacuum separations. This makes the dipole
correction a very useful tool for all problems where slabs with nonvanishing
dipole moments or inequivalent surface terminations have to be considered.

For the BaO--terminated BaTiO$_3$ slabs and the TiO$_2$--terminated
PbTiO$_3$ slabs as thin as 7 atomic layers, we have found strong evidence
for the presence of an instability to a spontaneously polarized ground
state with the polarization perpendicular to the surface. This is in
agreement with recent experiments \cite{tybell} and a microscopic
effective-Hamiltonian study.\cite{rabe} However, because the latter approach
does not explicitly take into account the structural relaxations at the
surface, it cannot provide the detailed level of description that our theory
gives. In particular, we find an enhancement of the polarization at the
surface for which the polarization points inwards, whereas the polarization
is reduced when the spontaneous polarization points outwards. We have also
detailed the variations in these surface relaxations as the polarization of
the slab is modified by an external electric field.

In order to obtain an even more realistic level of description, the next
obvious step will be to calculate the influence of a real metal/perovskite
interface on the atomic relaxations and the polarization at the interface.
However, such a study requires the choice of a particular metallic overlayer
with particular epitaxy conditions, and is thus intrinsically less universal.
Nevertheless, it may be an interesting avenue for future investigation.


\section{Acknowledgments}

We wish to thank Karin Rabe for useful discussions.
This work is supported by the ONR grant N00014-97-1-0048.



\end{document}